\documentclass[preprint,superscriptaddress,preprintnumbers,amsmath,amssymb,pra]{revtex4}
\usepackage{graphicx}
\usepackage{amsmath}%\usepackage{slashed}

\begin{document}
\newcommand{\ve}{\varepsilon}
\newcommand{\veD}{\varepsilon_D}
\newcommand{\veG}{\varepsilon_G}
\thispagestyle{empty}

\title{The Casimir force, causality and the Gurzhi model }

\author{
G.~L.~Klimchitskaya}
\affiliation{Central Astronomical Observatory at Pulkovo of the
Russian Academy of Sciences, Saint Petersburg,
196140, Russia}
\affiliation{Institute of Physics, Nanotechnology and
Telecommunications, Peter the Great Saint Petersburg
Polytechnic University, Saint Petersburg, 195251, Russia}

\author{
V.~M.~Mostepanenko}
\affiliation{Central Astronomical Observatory at Pulkovo of the
Russian Academy of Sciences, Saint Petersburg,
196140, Russia}
\affiliation{Institute of Physics, Nanotechnology and
Telecommunications, Peter the Great Saint Petersburg
Polytechnic University, Saint Petersburg, 195251, Russia}
\affiliation{Kazan Federal University, Kazan, 420008, Russia}
\author{Kailiang Yu}
\affiliation{Department of Physics, University of South Florida, Tampa, Florida 33620, USA}

\author{L.~M.~Woods}
\affiliation{Department of Physics, University of South Florida, Tampa, Florida 33620, USA}

\begin{abstract}
An extended Drude model, termed as the Gurzhi model, which takes into account
the electron-phonon and electron-electron interactions, is applied to calculate
the Casimir force between two metallic plates. It is shown that although the
dielectric permittivity of the Gurzhi model has a first order pole in the upper
half-plane of complex frequencies and, thus, violates the causality principle,
it can be used in a restricted frequency interval in combination with the
experimental permittivity determined by the optical data for the complex index
of refraction. The imaginary part of the Gurzhi dielectric permittivity of Au
at low frequencies demonstrates better agreement with the permittivity given
by the optical data than the simple Drude model. The Casimir pressure between
two Au plates is computed using the Gurzhi, Drude and plasma model approaches,
taking into account the optical data, as well as with the simple Drude and
plasma models. The contribution of the electron-electron scattering to the
Casimir pressure is estimated. Although a comparison with the measurement data
of two precise experiments show that the Gurzhi model does not resolve the
Casimir puzzle, the obtained results suggest further clarification of this
fundamental problem.
\end{abstract}

\maketitle

\section{Introduction}

It is common knowledge that the zero-point and thermal fluctuations of the
electromagnetic field are responsible for several interesting physical phenomena
which attract much experimental and theoretical attention in the last few years.
The best known example is the fluctuation-induced force acting between two
closely spaced uncharged bodies in vacuum. At  separations below
a few nanometers one can neglect the presence of retardation  and 
this force is known as the van der Waals force. At larger
separations, where the effects of retardation come into play,
it is conventional to speak about the Casimir force (see the recent reviews
\cite{1,2,3}). Theoretical description of the van der Waals and Casimir forces
is given by the Lifshitz theory \cite{4} which is derivable from the
fluctuation-dissipation theorem of quantum statistical physics, or from the
scattering approach, or by summing up the oscillator free energies in the framework
of quantum field theory \cite{5,5a,6,7}.

The Lifshitz theory allows computation of the van der Waals and Casimir energies,
free energies and forces given the frequency-dependent dielectric permittivities
of the interacting bodies are available. These permittivities are obtained from
the complex index of refraction which has been measured for a number of materials
over some frequency regions \cite{8}. The characteristic feature of
the van der Waals and Casimir forces is that their calculation requires a knowledge
of the dielectric permittivities over very wide frequency regions including at zero
frequency. The latter contributes significantly to the final result. Because of this,
it is necessary to extrapolate the available optical data down to zero frequency
on theoretical grounds. As an example, for metals extrapolations of this kind are
usually made by means of the Drude model which takes into account the
electron-phonon interaction at low frequencies.
The physical significance and properties of the Drude model are
discussed in detail in Ref.~\cite{8a}.

Surprisingly, in a number of experiments performed by two different
groups \cite{9,10,11,12,13,14,15,16,17,18,18a} it was found that the measurement results
exclude theoretical predictions of the Lifshitz theory obtained using an extrapolation
of the optical data by means of the Drude model. The same results turned out to be in
agreement with theory if the optical data reflecting the role of core (bound)
electrons are extrapolated down to zero frequency by the free-electron plasma
model \cite{9,10,11,12,13,14,15,16,17,18,18a}. An important role in these comparisons
is played by the contribution of the zero-frequency term of the Lifshitz formula
which depends heavily on the extrapolation used.
It seems quite unusual that the Lifshitz
theory is excluded by the measurement data if the actual relaxation properties of
conduction electrons at low frequencies are taken into account but it agrees with the
data if these properties are disregarded. Taking into consideration that the difference
between two alternative theoretical predictions at the experimental separations of
Refs.~\cite{9,10,11,12,13,14,15,16,17,18,18a} was below a few percent, attempts of
 solving the problem at the expense of some background effects or computational
 inaccuracies have been undertaken (see, e.g., Refs.~\cite{19,20,21,22}).

 The experimental situation was finally cleared up employing the differential force
 measurements proposed in Ref.~\cite{23} where the theoretical predictions of the
 Lifshitz theory obtained with the help of the Drude- and plasma-model extrapolations
 of the optical data differ by up to a factor of 1000. The experiment of this
 kind \cite{24} conclusively excluded an extrapolation by means of the Drude model
 and turned out to be in agreement with theoretical results using the plasma model.

A disagreement between theoretical predictions of the Lifshitz theory
obtained using the physically justified Drude model and
measurement results from many experiments is considered as puzzling 
\cite{25}. The roots of the Casimir puzzle are directly related to the fact
 that according to the Drude model
there is no contribution to the Casimir force from the transverse
electric mode at zero frequency \cite{25a}.
As a result, at large separations
between the interacting  plates the predicted Casimir force is only one half of
the one predicted using the plasma model. The single experiment
performed at large separations up to 7.3 micrometers \cite{25b} was 
interpreted as being in agreement with the Drude model prediction. In this 
experiment, however, not the Casimir force alone, but up to an order of 
magnitude larger force presumably originating from the so-called 
patch potentials was measured. The Casimir force itself was 
obtained indirectly using  large subtraction of some 
analytic expression containing two 
fitting parameters. According to Refs.~\cite{25c,25d}, this makes the 
results of Ref.~\cite{25b} uncertain. Various aspects of the Casimir 
puzzle are discussed at length in
 Refs.~\cite{1,3,7,8a,18a,24,25,25e,25f,25g}.

 Taking into account that a fundamental understanding of 
  the Casimir puzzle  is still missing, it seems warranted to reconsider
 the response of metals to the low-frequency electromagnetic field used in the
 Lifshitz theory. In the low frequency range, the electromagnetic response is
 determined by the intraband part which is essentially governed by the behavior
 of conduction electrons. Experimental studies using a variety of spectroscopic
 techniques show that much insight can be gained in the band structure properties
 of the materials as well as the scattering processes carriers exhibit in their
 dynamics \cite{26}. The relaxation parameter $\gamma_{ep}$ of the standard
 Drude model is determined by the electron-phonon scattering. However, at low
 frequencies electron-electron, electron-impurity, electron-surface and other
 interactions contribute to the total relaxation parameter as well (see
 Ref.~\cite{26a} and review \cite{27}), where in clean
 metallic systems electron-electron
 scattering is the major addition to the electron-phonon one.
 It has been known that for noble metals the contribution of electron-electron
 scattering $\gamma_{ee}$ to the relaxation parameter can be described by  the Gurzhi
 formula \cite{27,28,29,30} which contains both the frequency- and
 temperature-dependent terms. Replacing $\gamma_{ep}$ in the dielectric
 permittivity of the Drude model with $\gamma_{ep}+\gamma_{ee}$, one obtains the
 so-called extended Drude, or Gurzhi model for the dielectric permittivity of metals.

 In this paper, we investigate possible applications of the Gurzhi model in the
 Lifshitz theory for calculations of the Casimir force. We explore the analytic
 properties of the Gurzhi dielectric permittivity and demonstrate that it violates
 the causality condition which precludes its use over the entire frequency axis.
 Next, we consider the dielectric permittivities of the Gurzhi, Drude and plasma
 models in combination with the measured optical data. It is confirmed that over some
 frequency region below 2~eV the Gurzhi model provides a better analytic
 approximation to the measured imaginary part of the dielectric permittivity of
 Au than the Drude model. The Casimir pressure between two parallel plates made
 of Au is computed using the optical data extrapolated down to zero frequency
 by means of the Gurzhi, Drude and plasma models, and the obtained results are
 compared. This allowed estimation of the possible role of
 electron-electron interactions
 in the Casimir force. The Casimir pressures computed with different models of the
 dielectric permittivity are correlated with precise experiments on measuring
 the Casimir interaction. It is shown that although the Gurzhi model provides
 a better analytic approximation to the optical data in some frequency range
 than the Drude one, it
 does not resolve the Casimir puzzle.

 The paper is organized as follows. In Sec.~II we describe the main features of
 the Gurzhi model and consider its analytic properties in connection with the
 causality principle. Section~III is devoted to comparison between different
 analytic models of the dielectric permittivity of Au combined with the measured
 optical data. In Sec.~IV the Casimir interaction computed using different
 permittivities is compared with the measurement results of two precise
 experiments. In Sec.~V the reader will find our conclusions and discussion.

\section{The Gurzhi dielectric permittivity and its analytic properties}

It is well known that at sufficiently low frequencies the response of metals to
electromagnetic field is essentially described by the dielectric permittivity
of the Drude model
\begin{equation}
\veD(\omega,T)=1-
\frac{\omega_p^2}{\omega[\omega+i\gamma_{ep}(T)]},
\label{eq1}
\end{equation}
\noindent
where $\omega_p$ is the plasma frequency and $\gamma_{ep}(T)$ is the
temperature-dependent relaxation parameter determined by the process of
electron-phonon scattering.

An extended version of the Drude dielectric permittivity, which is called sometimes the Gurzhi model, takes a similar form
\begin{equation}
\veG(\omega,T)=1-
\frac{\omega_p^2}{\omega[\omega+i\gamma(\omega,T)]}.
\label{eq2}
\end{equation}
\noindent
Here, however, the relaxation parameter consists of two terms
\begin{equation}
\gamma(\omega,T)=\gamma_{ep}(T)+\gamma_{ee}(\omega,T)
\label{eq3}
\end{equation}
\noindent
taking into account the processes of electron-phonon and electron-electron
scattering.

The theoretical expression for $\gamma_{ee}$ was derived in Ref.~\cite{28}
(see also Refs.~\cite{27,29,30}) based on the Boltzmann quantum equation
for the electronic Fermi liquid and the Kubo formula which relates the
conductivity to the current-current correlation function
\begin{equation}
\gamma_{ee}(\omega,T)=D\left[(k_BT)^2+
\left(\frac{\hbar\omega}{2\pi}\right)^2\right].
\label{eq4}
\end{equation}
\noindent
Here, the coefficient $D=\pi^3\Gamma\Delta/(12\hbar E_F)$ where $E_F$ is
the Fermi level of the metal under consideration, $k_B$ is the Boltzmann
constant, $\Delta=0.75$  is the fractional umklapp scattering, and
$\Gamma=0.55$ is the averaged scattering probability over the Fermi surface.
Note that Eq.~(\ref{eq4}) has been verified in several experiments
for noble metals
\cite{31,32,33,34} in the near infrared frequency range up to the
interband absorption frequencies. In so doing, the temperature-dependent
contribution in Eq.~(\ref{eq4}) is small as compared to the
frequency-dependent one. For Au, which is the metal of our interest below,
one has $D=0.94~\mbox{fs}^{-1}\mbox{eV}^{-2}$ \cite{30}.

Substituting Eq.~(\ref{eq4}) in Eq.~(\ref{eq3}), the relaxation parameter
taking into account both the electron-phonon and electron-electron scattering
can be written in the form
\begin{equation}
\gamma(\omega,T)=C(T)+B\omega^2,
\label{eq5}
\end{equation}
\noindent
where
\begin{equation}
C(T)=\gamma_{ep}(T)+D(k_BT)^2, \quad
B=D\left(\frac{\hbar}{2\pi}\right)^2.
\label{eq6}
\end{equation}

The dielectric permittivity of the Gurzhi model (\ref{eq2}), (\ref{eq5}),
besides the singular point at $\omega=0$, has poles in the plane of complex
frequencies determined by the roots of the quantity
\begin{equation}
iB\omega^2+\omega+iC(T)=0.
\label{eq7}
\end{equation}
\noindent
By solving this equation, one obtains
\begin{equation}
\omega^{(1,2)}=i\xi^{(1,2)}=\frac{i}{2B}\left[1\pm\sqrt{1+4BC(T)}\right],
\label{eq8}
\end{equation}
\noindent
where $\omega^{(1)}$ and $\omega^{(2)}$ belong to the upper and lower half-planes,
respectively.

\begin{figure}[b]
\vspace*{-0cm}
\centerline{\hspace*{0.3cm}
\includegraphics{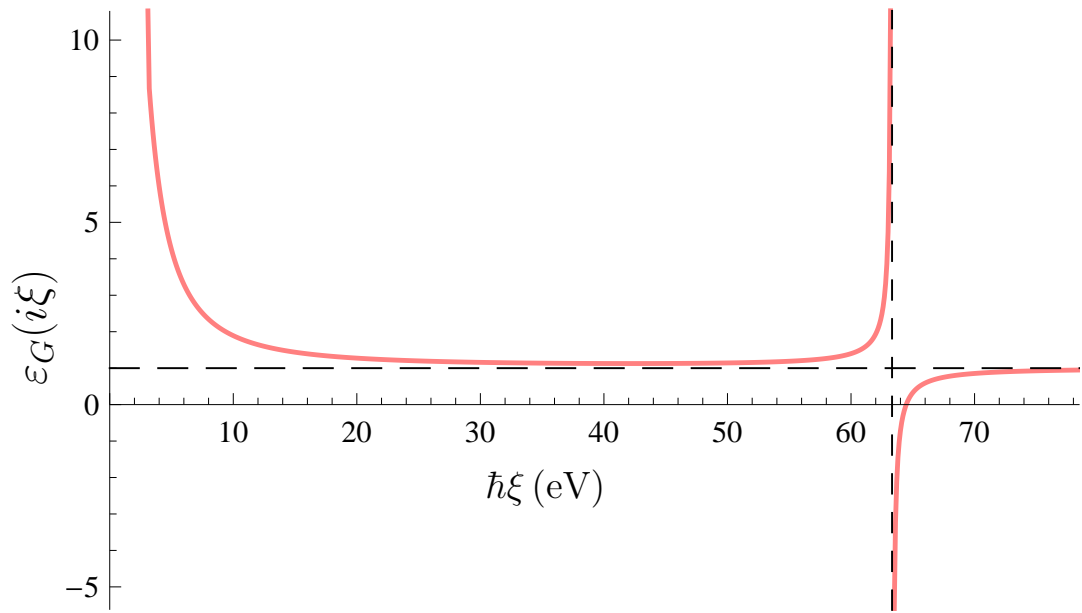}
}
\vspace*{-17.cm}
\caption{\label{fg1}
The dielectric permittivity of the Gurzhi model is shown as a
function of the imaginary frequency.
}
\end{figure}

Along the imaginary frequency axis $\omega=i\xi$ the Gurzhi dielectric permittivity
(\ref{eq2}), (\ref{eq4}) takes the real values
\begin{equation}
\veG(i\xi)=1+\frac{\omega_p^2}{\xi\left[\xi+C(T)-B\xi^2\right]}.
\label{eq9}
\end{equation}
\noindent
As an example, in Fig.~\ref{fg1} $\veG$ is shown as a function of $\xi$ for Au
at room temperature using the parameters of the Gurzhi model indicated above and
the experimental parameters of the Drude model $\hbar\omega_p=8.68$~eV and
$\hbar\gamma_{ep}(T=295\,\mbox{K})=30.3$~meV \cite{30}. From Eq.~(\ref{eq9}) and
Fig.~\ref{fg1} it is seen that the dielectric permittivity $\veG$ reaches the
minimum value $\veG(i\xi^{(m)})=1.1267$ at the point
\begin{equation}
\hbar\xi^{(m)}=\frac{\hbar}{3B}\left[1+\sqrt{1+3BC(T)}\right]=
42.2094\,\mbox{eV}
\label{eq10}
\end{equation}
\noindent
and has a break of continuity at the point $\hbar\xi^{(1)}=63.3217\,\,$eV defined
in Eq.~(\ref{eq8}). It is important to note that in the region from $\hbar\xi^{(1)}$
to $\hbar\xi^{(0)}\approx64.47$~eV the Gurzhi permittivity takes the negative values
and vanishes at $\xi=\xi^{(0)}$: $\veG(i\xi^{(0)})=0$. For $\xi>\xi^{(0)}$ $\veG$ increases
 monotonously and goes to unity when $\xi$ goes to infinity.

These properties are not normal for commonly employed dielectric permittivities
which must meet some necessary physical conditions. It has been known that the
electric displacement {\boldmath$D$}$(t)$ is determined by the values of electric
field {\boldmath$E$}$(t)$ at all {\it previous} moments of time \cite{35}
\begin{equation}
\mbox{\boldmath$D$}(t)=\mbox{\boldmath$E$}(t)+
\int_0^{\infty}f(\tau)\mbox{\boldmath$E$}(t-\tau)d\tau,
\label{eq11}
\end{equation}
\noindent
where the function of time $f(\tau)$ is finite at all $\tau$, depends on the
properties of a medium and defines the frequency-dependent dielectric permittivity
\begin{equation}
\ve(\omega)=1+
\int_0^{\infty}f(\tau)e^{i\omega\tau}d\tau.
\label{eq12}
\end{equation}
\noindent
Equations (\ref{eq11}) and (\ref{eq12}) constitute the mathematical
formulation of the principle of causality stating that future has no effect on the
past. From Eq.~(\ref{eq12}) it is seen that in the upper half-plane
(${\rm Im}\omega>0$) the integral converges and, thus, $\ve(\omega)$ has no
singular points \cite{35}. This statement is a consequence of the principle of
causality. It is easily seen also that in the upper half-plane (including the
real frequency axis) the dielectric permittivity cannot turn into zero \cite{35}.
All the above results in the Kramers-Kronig relations which link to one another
the real and imaginary parts of the dielectric permittivity.

{}From Fig.~\ref{fg1} and related discussion it is seen that the dielectric
permittivity of the Gurzhi model does not satisfy the principle of causality and the
Kramers-Kronig relations. Several attempts to recover the Kramers-Kronig relations
have been undertaken (see, for instance, Refs.~\cite{26,31,36}) by introducing the
so-called {\it memory function}, but its explicit form remains unavailable.
In fact to cancel the first order pole at the point $i\xi^{(1)}$ in the upper half-plane
of complex frequencies it would be necessary to replace the plasma frequency squared
in Eq.~(\ref{eq9}) with the frequency- and temperature-dependent quantity
\begin{equation}
{\tilde{\omega}}_p^2(\omega,T)=\left[2iB\omega+1+\sqrt{1+4BC(T)}\right]
g(\omega)\omega_p^2,
\label{eq13}
\end{equation}
\noindent
where $g(\omega)$ is any analytic function in the upper half-plane.
In this case, however, the physically meaningful term of the order of
$\omega^2$ in Eq.~(\ref{eq4}) would be lost.

Thus, one can conclude that the Gurzhi model can be used only in some restricted
region of low frequencies as more or less good phenomenological description for the
dielectric permittivity of noble metals. In this respect it would be interesting
to compare it with other analytic models used in computations of the Casimir
force and with the experimental permittivity obtained from the measured optical data.

\section{Different models of the dielectric permittivity and
the optical data}

It is well known that the Lifshits formulas for the Casimir free energy and
pressure are most conveniently expressed via the dielectric permittivity of plate
materials along the imaginary frequency axis. The latter quantity, in its turn,
can be found by means of the Kramers-Kronig relations
\begin{equation}
\ve(i\xi)=1+\frac{2}{\pi}\int_0^{\infty}
\frac{\omega\,{\rm Im}\ve(\omega)}{\omega^2+\xi^2}d\omega
\label{eq14}
\end{equation}
\noindent
or
\begin{equation}
\ve(i\xi)=1+\frac{2}{\pi}\int_0^{\infty}
\frac{\omega\,{\rm Im}\ve(\omega)}{\omega^2+\xi^2}d\omega
+\frac{\omega_p^2}{\xi^2}
\label{eq15}
\end{equation}
\noindent
expressing $\ve(i\xi)$ through the imaginary part of $\ve$ defined along the real
frequency axis. Equation (\ref{eq14})  is valid for the permittivities
that are regular
at zero frequency or have a first order pole \cite{7,35}, whereas Eq.~(\ref{eq15})
is obeyed by the permittivities having a second order pole and the residue
$\omega_p^2$ at zero frequency \cite{7,37}.

Here we compare the imaginary parts of the dielectric permittivities of Au found
from the measured optical data \cite{8} and given by the Drude and Gurzhi models
in the frequency region below 2~eV, i.e., well below the first absorption frequency.

\begin{figure}[b]
\vspace*{-5cm}
\centerline{\hspace*{0.3cm}
\includegraphics{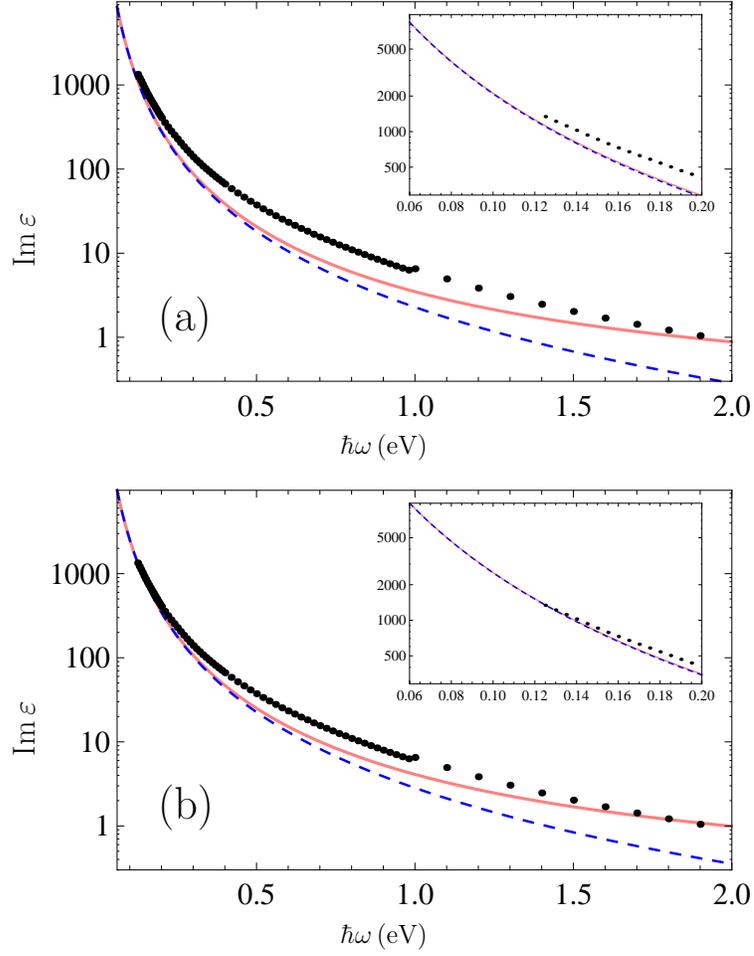}
}
\vspace*{-12.cm}
\caption{\label{fg2}
The imaginary part of the dielectric permittivity of Au is shown
as a function of frequency by the dots obtained from the optical data
for the complex index of refraction and by the solid and dashed lines
obtained using the Gurzhi and Drude models, respectively, with the plasma
frequency and relaxation parameter (a) $\hbar\omega_p = 8.68$~eV,
$\hbar\gamma_{ep} = 30.3$~meV and (b) $\hbar\omega_p = 9.0 $~eV,
$\hbar\gamma_{ep} = 35.0$~meV. The region of small frequencies is shown
in the insets on an enlarged scale.
}
\end{figure}

In Figs.~\ref{fg2}(a) and \ref{fg2}(b) the imaginary part of the dielectric
permittivity of Au is shown by dots using the values for real and imaginary parts
of the complex index of refraction of Au measured at frequencies above 0.125~eV
\cite{8}. The solid and dashed lines demonstrate the imaginary part of the dielectric
permittivity of Au given by the Gurzhi (\ref{eq2}) and Drude (\ref{eq1}) models,
respectively.  In Fig.~\ref{fg2}(a) the experimental values of parameters  $\hbar\omega_p=8.68$~eV and
$\hbar\gamma_{ep}(T=295\,\mbox{K})=30.3$~meV \cite{30} have been used.
These parameters, however, are sample-dependent \cite{19}.
Because of this, in Fig.~\ref{fg2}(b) ${\rm Im}\ve$ given by the Gurzhi and
Drude models are plotted using the values $\hbar\omega_p=9.0$~eV and
$\hbar\gamma_{ep}(T=295\,\mbox{K})=35.0$~meV which were found most appropriate
for Au films employed in precise measurements of the Casimir force
\cite{9,10,11,12,13,14,15,16,17,18,18a,24}.

As is seen in both Figs.~\ref{fg2}(a) and \ref{fg2}(b), the Gurzhi model better
reproduces the optical data than the Drude model over the frequency region from
0.3 to 2~eV. From the comparison of Fig.~\ref{fg2}(a) and Fig.~\ref{fg2}(b)
it is seen also that the values of $\omega_p$ and $\gamma$ used in experiments
on measuring the Casimir force result in a better agreement between ${\rm Im}\ve$
obtained from the optical data and from  the Gurzhi and
Drude models than the values of Ref.~\cite{30}. This result is readily illustrated
by comparing insets in  Figs.~\ref{fg2}(a) and \ref{fg2}(b) where the frequency
regions from 0.06 to 0.2~eV are shown on an enlarged scale.
{}From the inset in  Fig.~\ref{fg2}(a) one can see that there is a break of
continuity between the values of ${\rm Im}\ve$ found from the optical data at the
minimum frequency, where they are available, and from the analytic models.
By contrast, in the inset to  Fig.~\ref{fg2}(b) there is a smooth extrapolation
of the optical data by the Gurzhi and
Drude models in the frequency region from 0 to 0.125~eV.

As mentioned in Sec.~I, one of the approaches to the calculation of the Casimir force
consists in using the optical data for ${\rm Im}\ve$ over the entire frequency range
where they are available (from 0.125 to 9919~eV), extrapolated to below 0.125~eV 
(i.e., in the region from 0 to 0.125~eV)
by means of the imaginary part of the Drude model [i.e., by the dashed line in
Fig.~\ref{fg2}(b)]. In doing so the values of $\ve(i\xi)$ are obtained from
Eq.~(\ref{eq14}) (there is no need in extrapolation of the optical data
to the region above 9919~eV).
This is the so-called Drude model approach to calculating the Casimir
force which takes into account all real processes involving conduction and core
electrons at frequencies above 0.125~eV and the electron-phonon interaction occurring
at lower frequencies.

Another approach to calculate the Casimir force uses ${\rm Im}\ve$ determined by
the optical data of Au only over the frequency range from 2 to 9919~eV related to
interband transitions. It is assumed that ${\rm Im}\ve=0$ within the frequency
region from 0 to 2~eV, i.e., all the processes involving conduction electrons are
disregarded. It is assumed also that the dielectric permittivity has a pole of
 second order and a residue equal to $\omega_p^2$ at zero frequency. Then,
the dielectric permittivity along the imaginary frequency axis is found from
Eq.~(\ref{eq15}). This is called the plasma model approach to the calculation of the
Casimir force. As described in Sec.~I, the Casimir puzzle lies in the fact that
all precise experiments on measuring the Casimir force at separations below
$1~\mu$m exclude the Drude model approach
considered, and are in good agreement with the
plasma model one (see also Sec.~IV).

In the frequency region, where the Gurzhi dielectric permittivity is in reasonably
good agreement with the optical data (i.e., below 2~eV), it could also be applied
for calculating the Casimir force. For this purpose,  at $\hbar\omega>2$~eV
${\rm Im}\ve$ is obtained
from the optical data and at $\hbar\omega<2$~eV
it is given by the imaginary part of
the Gurzhi model. Then the dielectric permittivity along
the entire imaginary frequency axis is found from Eq.~(\ref{eq14}). This could be called
the Gurzhi model approach to the Casimir force.
In the frequency region from 0.125 to 2~eV it takes into
account analytically both the electron-phonon and electron-electron interactions
(taken into account via the optical data in the Drude model approach). Both these
processes are accounted for also at $\hbar\omega<0.125$~eV [see the solid line in
Fig.~\ref{fg2}(b)], whereas the Drude model approach disregards the electron-electron
scattering within this frequency region.

In the end of this section, we  particularly emphasize that the type of
singularity of $\ve$ at zero frequency makes a profound effect on its values at 
pure imaginary frequencies. As shown in Ref.~\cite{38}, the behavior of $\ve(i\xi)$
over the entire axis $0<\xi<\infty$ can be found theoretically
using the available optical data for the complex index of refraction with no
any extrapolation. This is achieved through the
application of the so-called weighted Kramers-Kronig transform, which suppresses a
contribution of the frequency regions where the optical data are not available, and
assumes the presence of either the first or the second order pole of $\ve$ at zero frequency.

\section{Calculation of the Casimir pressure in different approaches and comparison
with experiments}

The Casimir pressure between two parallel metallic plates of more than 100~nm
thickness at temperature $T$ separated by the vacuum gap of width $a$ is the same
as between two semispaces. It is given by the Lifshitz formula \cite{1,2,3,4,5,7}
\begin{eqnarray}
&&
P(a,T)=-\frac{k_BT}{\pi}\sum_{l=0}^{\infty}{\vphantom{\sum}}^{\prime}
\int_0^{\infty}\!\!\!q_lk_{\bot}dk_{\bot}
\nonumber \\
&&~~~~~~~\times
\sum_{\alpha}
\left[r_{\alpha}^{-2}(i\xi_l,k_{\bot})e^{2aq_l}-1\right]^{-1},
\label{eq16}
\end{eqnarray}
\noindent
where the prime on the first summation sign
corresponds to dividing the term with $l=0$ by 2,
$k_{\bot}$ is the magnitude of projection of the wave vector on the plane of plates,
$\alpha$ implies a summation over the transverse magnetic ($\alpha={\rm TM}$) and
transverse  electric ($\alpha={\rm TE}$) polarizations of the electromagnetic field,
$\xi_l=2\pi k_BTl/\hbar$ with $l=0,\,1,\,2,\,\ldots$ are the Matsubara frequencies, and
$q_l=(k_{\bot}^2+{\xi_l^2}/{c^2})^{1/2}$.

The reflection coefficients in Eq.~(\ref{eq16}) are defined as
\begin{equation}
r_{\rm TM}(i\xi_l,k_{\bot})=\frac{\ve_lq_l-k_l}{\ve_lq_l+k_l},
\quad
r_{\rm TE}(i\xi_l,k_{\bot})=\frac{q_l-k_l}{q_l+k_l},
\label{eq18}
\end{equation}
\noindent
\noindent
where
\begin{equation}
\ve_l\equiv\ve(i\xi_l), \quad
k_l=\left(k_{\bot}^2+\ve_l\frac{\xi_l^2}{c^2}\right)^{1/2}.
\label{eq19}
\end{equation}
\noindent
{}From Eq.~(\ref{eq4}) it is seen that at the first Matsubara frequency it holds
$\gamma_{ee}(i\xi_1,T)=0$. This is the so-called
{\it first-Matsubara-frequency} rule \cite{27}.

\begin{figure}[b]
\vspace*{-4cm}
\centerline{\hspace*{0.3cm}
\includegraphics{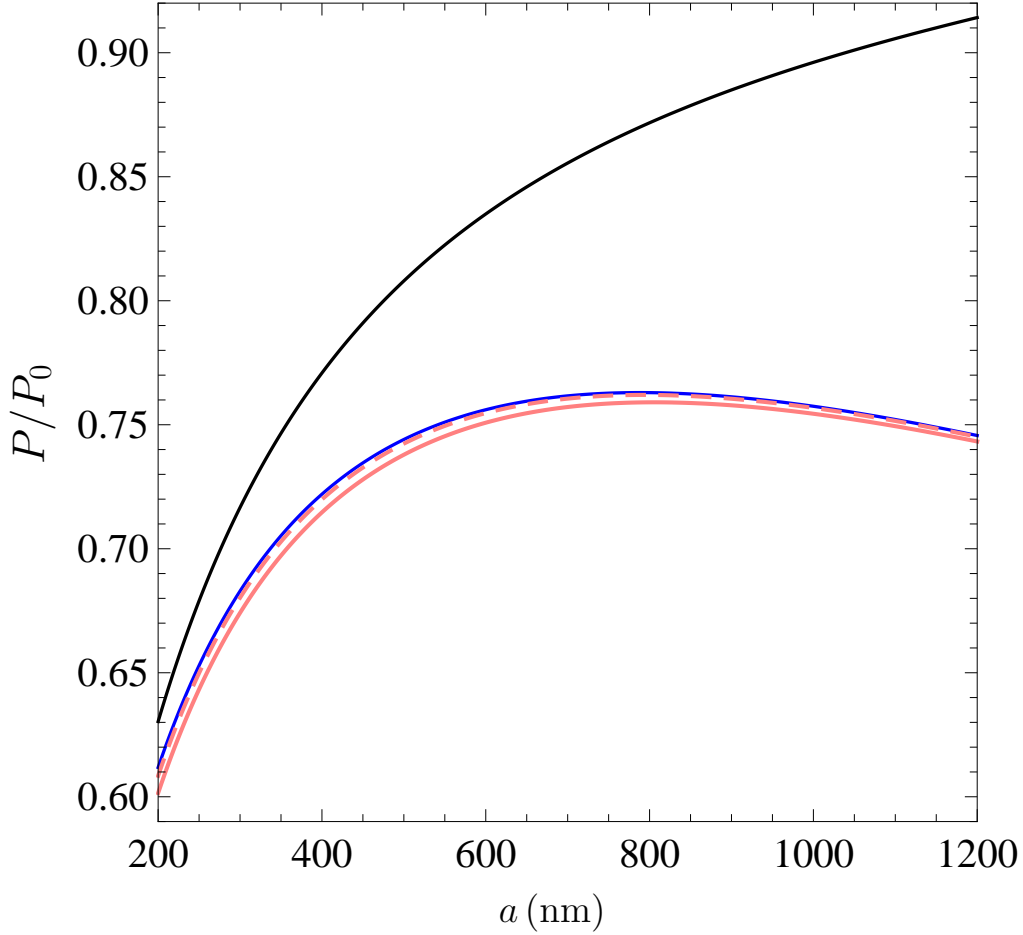}
}
\vspace*{-13.cm}
\caption{\label{fg3}
 The ratio of the Casimir pressure between two parallel Au plates
obtained at $T=295~$K using an extrapolation of the optical data in the region from 0 to 2~eV by
means of the plasma model with $\hbar\omega_p = 9.0 $~eV (the top solid line),
in the region from 0 to
0.125~eV by means of the Drude model with $\hbar\omega_p = 9.0 $~eV,
$\hbar\gamma_{ep} = 35.0$~meV (the middle solid line), and 
in the region from 0 to  2~eV by means of the Gurzhi
model with $\hbar\omega_p = 8.68$~eV, $\hbar\gamma_{ep} = 30.3$~meV 
(the bottom solid line)
to the pressure between two ideal metal plates is shown
as function of separation. The dashed line shows the
same ratio where the numerator is calculated using  the Gurzhi model with
$\hbar\omega_p = 9.0 $~eV and $\hbar\gamma_{ep} = 35.0$~meV.
}
\end{figure}

We have calculated the ratio of the Casimir pressure (\ref{eq16}) at $T=295~$K to
that between two ideal metal plates at zero temperature,
$P_{0}(a)=-{\pi^2\hbar c}/(240a^4)$,
as a function of separation between the plates, using three theoretical approaches
described in the end of Sec.~III, i.e., the plasma, Drude,  and Gurzhi.
The computational results are presented in Fig.~\ref{fg3} by the three solid lines
counted from top to bottom. These lines are obtained by using extrapolations of
the optical data to below 2~eV by means of the plasma model with
$\hbar\omega_p=9.0$~eV, to below 0.125~eV by means of the Drude model with
$\hbar\omega_p=9.0$~eV, $\hbar\gamma_{ep}=35$~meV, and to below 2~eV by means
of the Gurzhi model with
$\hbar\omega_p=8.68$~eV, $\hbar\gamma_{ep}(T=295\,\mbox{K})=30.3$~meV,
respectively.  The dashed line is found by extrapolating the optical data
to below 2~eV using the Gurzhi model with
$\hbar\omega_p=9.0$~eV and $\hbar\gamma_{ep}=35$~meV.

As is seen in Fig.~\ref{fg3}, the Drude and Gurzhi approaches
lead to rather close results for the Casimir pressure, especially if the Gurzhi model
employs the same Drude parameters as the Drude model (see the dashed line).
At the same time, the computational results found within the plasma model approach
are quite different. This is explained by distinct behaviors of the
dielectric permittivities at zero frequency (the first order pole in the cases of
Drude and Gurzhi models and the second order one for the plasma model).

To quantify the role of the optical data below 2~eV and at higher frequencies in the
region of absorption bands, in Table~I we present several computational results
found with the partial or total exclusion of the use
of optical data in favor of the
simple plasma or Drude models. Columns 3, 5, and 8 of Table~I contain magnitudes
of the Casimir pressure computed using the plasma, Drude and Gurzhi model
approaches, respectively, at separation distances indicated in column~1.
These computations are performed with the optical data, as described in explanations to
Fig.~\ref{fg3}, with the Drude parameters
$\hbar\omega_p=9.0$~eV and $\hbar\gamma_{ep}=35$~meV.
In column 2 of Table~I, we present the mean measured magnitudes of the Casimir
pressure and their total experimental errors determined at the 95\% confidence level
by the results of Refs.~\cite{11,12}. In columns 4 and 7, the magnitudes of the
Casimir pressure computed by using the simple plasma and Drude models are presented,
respectively, i.e.,
\begin{equation}
\ve_p(i\xi_l)=1+\frac{\omega_p^2}{\xi_l^2}, \quad
\veD(i\xi_l)=1+\frac{\omega_p^2}{\xi_l[\xi_l+\gamma_{ep}(T)]}.
\label{eq21}
\end{equation}
\noindent
Finally, column 6 contains the computational results using the optical data of Au
at $\hbar\omega>2$~eV and the simple Drude model at $\hbar\omega\leq 2$~eV.

We note that computations of the Casimir pressure by using the simple Gurzhi model
(\ref{eq2}) applied over the entire frequency range
would be inconsistent. The reason is that in the region from
$\hbar\xi^{(1)}=63.3217$~eV to $\hbar\xi^{(0)}=64.47$~eV one has
$\veG(i\xi)<0$ (see Sec.~II). The width of the frequency interval where $\veG(i\xi)$
is negative is almost independent of the values of the Drude parameters.
For instance, for $\hbar\omega_p=9.0$~eV and $\hbar\gamma_{ep}=35$~meV it holds
$\hbar\xi^{(1)}=63.3261$~eV in place of $63.3217$~eV. As a result, at room
temperature ($T=295~$K) one obtains that  $\veG(i\xi_l)$ with $397\leq l\leq 403$
takes the negative values. This leads to the complex $k_l$ in Eq.~(\ref{eq19})
within some interval of $k_{\bot}$ and, finally, to the complex reflection
coefficients and Casimir pressures in Eq.~(\ref{eq16}).

It may be argued that the Matsubara terms with such high $l$ do not contribute to the
pressure at separations considered. The presence of the complex-valued terms
(even though they were negligibly small in magnitudes) is, however, quite
impermissible theoretically. Furthermore, at sufficiently short separations between
the plates it is necessary to take into account much larger number of Matsubara
terms in order to calculate the Casimir pressure with sufficient precision.
Usually one should include all terms up to $15\omega_c=15c/(2a)$ \cite{7}.
As a result, at $a=15~$nm the first 650 Matsubara terms should be included  at room
temperature. This makes apparent that the simple Gurzhi model cannot be used
over the entire frequency range not
only theoretically but from the practical standpoint as well.

Now we discuss a correlation between the magnitudes of the Casimir pressure in
columns 3--8 of Table~I. We note that these values are burdened by the errors of
approximately 0.5\% determined by inaccuracies in the optical data and values
of parameters in the models used \cite{7}. An interrelationship between the values in
columns 3, 5, and 8 (obtained using the plasma, Drude, and Gurzhi approaches,
respectively) is the same as already discussed above for respective lines in
Fig.~\ref{fg3}. By comparing column 4 with column 3, it is seen that the
use of the simple plasma model (column 4) results in slightly smaller
magnitudes of the Casimir pressure,
and the impact of the optical data becomes more pronounced with decreasing
separation between the plates.

If one uses the simple Drude model at all frequencies below 2~eV
(column 6), slightly smaller magnitudes of the Casimir pressure are obtained as
compared to column 5, where the optical data are extrapolated down to zero
frequency by the simple Drude model in the region $\hbar\omega<0.125$~eV.
When the simple Drude model is applied over the entire frequency axis (column 7),
even smaller values for the magnitudes of the Casimir pressure are obtained.
With increasing separation, however, differences between the Casimir pressures
in columns 5, 6, and 7 become negligibly small which reflects a decreasing impact
of the optical data in the region of absorption bands on the computational results.
Note also that the Gurzhi model approach to calculation of the Casimir force
(column 8) leads to almost the same (but slightly larger) pressure magnitudes
as those in column 6 obtained using the simple Drude model below 2~eV and the
optical data at all higher frequencies.

A comparison between the Casimir pressures found with the Gurzhi model approach
(column 8) and by means of the Drude model below 2~eV (column 6) allows estimation
of the role of electron-electron scattering in the Casimir interaction.
At $a=0.2~\mu$m it contributes only about 0.16\% of the pressure and
its contribution decreases
with increasing separation.

By comparing the experimental Casimir pressures in column 2 with the theoretical ones
 in columns 3--8 one can conclude that in the limits of experimental and
theoretical errors the
measurement data are in agreement with the theoretical predictions made using the plasma
model approach and exclude the predictions of all other approaches.
This conclusion can be made quantitative taking into account that all precise
experiments on measuring the Casimir interaction have been performed in the
sphere-plate geometry (rather than in the plate-plate one) and that the test bodies have
some surface roughness, which is not taken into account in the theoretical results of
columns 3--8.

In the experiment of Refs.~\cite{11,12}, performed by means of a micromechanical
oscillator, the immediately measured quantity was the gradient of the Casimir force
acting between the sphere of $R=150~\mu$m radius and  a plate. This quantity
can be recalculated in the magnitude of the Casimir pressure between two parallel
plates presented in column 2 of Table~I using the proximity force approximation
\cite{1,7}
\begin{equation}
|P(a,T)|=\frac{1}{2\pi R}\,\frac{\partial F_{sp}(a,T)}{\partial a}.
\label{eq22}
\end{equation}
\noindent
The relative corrections to an approximate expression (\ref{eq22}),
which are less than $a/R$ \cite{21,22},
are negligibly small in this experiment. Small corrections due to the surface roughness
have been taken into account perturbatively \cite{1,7,39} in the Casimir pressure
$P_{\rm theor}$ (note that the surface roughness plays a more important role at very
short separations between the test bodies \cite{40}).

\begin{figure}[b]
\vspace*{-2cm}
\centerline{\hspace*{0.3cm}
\includegraphics{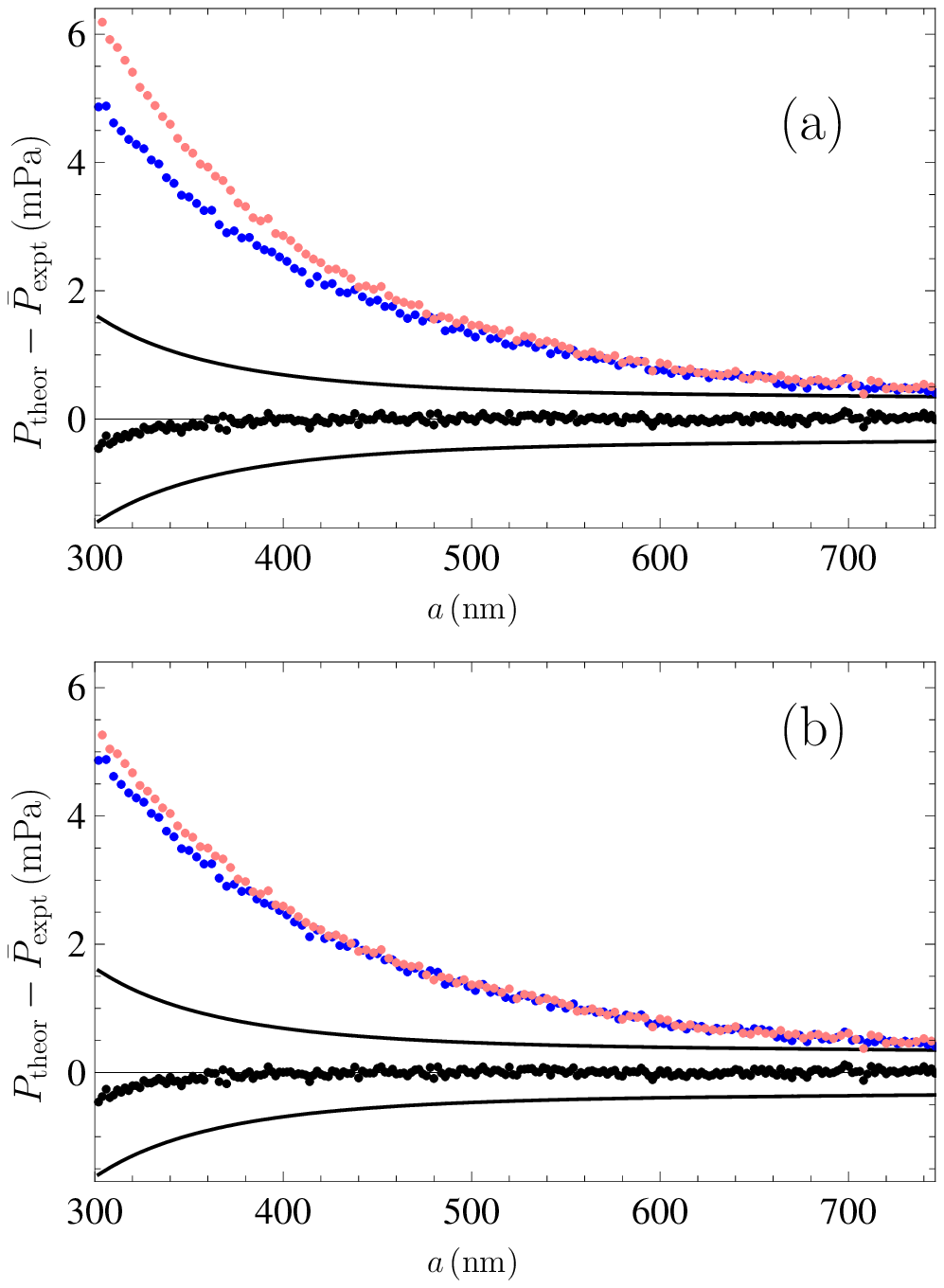}
}
\vspace*{-13.cm}
\caption{\label{fg4}
The differences between theoretical and mean experimental
\cite{11,12} Casimir pressures at $T=295~$K computed using an extrapolation
of the optical data of Au in the region from 0 to  2 eV by means of the plasma model
with $\hbar\omega_p = 9.0 $~eV (the bottom set of dots),
in the region from 0 to 0.125~eV by means of the Drude
model with $\hbar\omega_p = 9.0$~eV, $\hbar\gamma_{ep} = 35.0$~meV
(the middle set of dots),
and in the region from 0 to  2~eV by means of the Gurzhi model 
(the top set of dots) are shown as the functions of separation.
The Drude parameters in the Gurzhi model are (a) $\hbar\omega_p = 8.68$~eV,
$\hbar\gamma_{ep} = 30.3$~meV and (b) $\hbar\omega_p = 9.0$~eV,
$\hbar\gamma_{ep} = 35.0$~meV.
Two solid lines indicate the borders of the confidence intervals found
at the 95\% confidence probability.
}
\end{figure}

In Fig.~\ref{fg4}, we plot the differences between the theoretical Casimir pressures
computed using the plasma, Drude and Gurzhi model approaches and mean experimental
pressures measured in Refs.~\cite{11,12} (three sets of dots counted from bottom to
top, respectively) as the functions of separation. The Drude parameters in the Gurzhi
model are chosen as
(a) $\hbar\omega_p=8.68$~eV and $\hbar\gamma_{ep}=30.3$~meV and
(b) $\hbar\omega_p=9.0$~eV and $\hbar\gamma_{ep}=35$~meV.
The solid lines are formed by the borders of the confidence intervals found at each
separation by combining the total experimental and theoretical errors determined at
the 95\% confidence probability. As is seen in Figs.~\ref{fg4}(a) and \ref{fg4}(b),
both the Drude model approach and the Gurzhi model one used with any set of
the Drude
parameters are excluded by the measurement data at the 95\% confidence level,
whereas the plasma model approach is experimentally consistent.

We also compare the theoretical predictions of all three approaches with the recently
measured gradient of the Casimir force acting between the Au-coated surfaces of a
sphere and a plate refined by means of the UV and Ar-ion cleaning \cite{18}.
In this work, performed by means of an atomic force microscope, the sphere radius
was reduced to $R=43~\mu$m, and the corrections due to the use of the proximity
force approximation have been taken into account through the results of Ref.~\cite{22}.
The corrections due to the surface roughness were also included in the theoretical
gradients of the Casimir force \cite{18}.

\begin{figure}[b]
\vspace*{-2cm}
\centerline{\hspace*{0.3cm}
\includegraphics{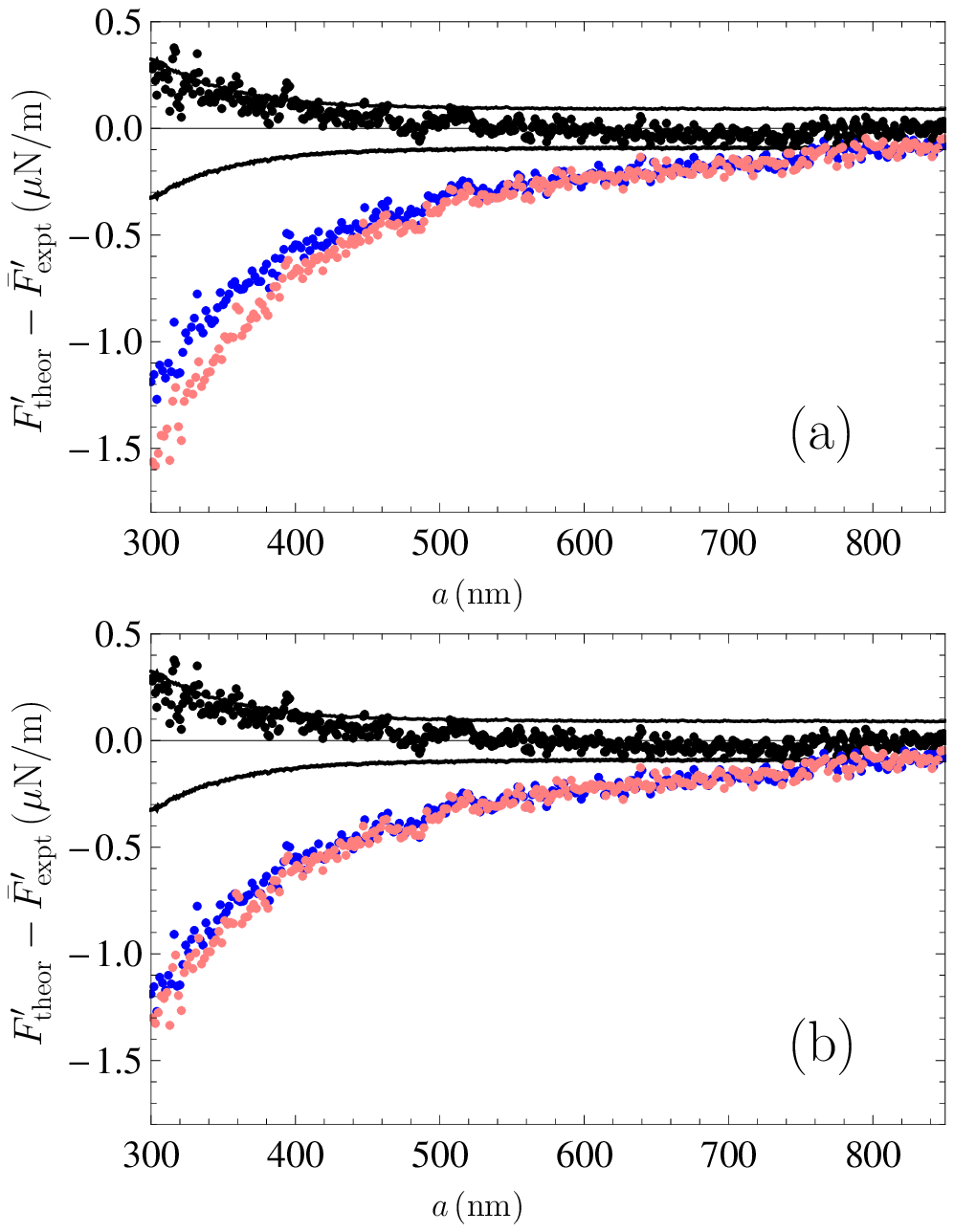}
}
\vspace*{-13.cm}
\caption{\label{fg5}
The differences between theoretical and mean experimental
\cite{18} gradients of the Casimir force at $T=295~$K
 computed using an extrapolation
of the optical data of Au in the region from 0 to  2 eV by means of the plasma model
with $\hbar\omega_p = 9.0 $~eV (the top set of dots),
in the region from 0 to  0.125~eV by means of the Drude
model with $\hbar\omega_p = 9.0$~eV, $\hbar\gamma_{ep} = 35.0$~meV 
(the middle set of dots),
and in the region from 0 to  2~eV by means of the Gurzhi model (the bottom set of dots)
 are shown as the functions of separation.
The Drude parameters in the Gurzhi model are (a) $\hbar\omega_p = 8.68$~eV,
$\hbar\gamma_{ep} = 30.3$~meV and (b) $\hbar\omega_p = 9.0$~eV,
$\hbar\gamma_{ep} = 35.0$~meV.
Two solid lines indicate the borders of the confidence intervals found
at the 67\% confidence probability.
}
\end{figure}

In Fig.~\ref{fg5} the differences between the theoretical gradients of
the Casimir force
computed within the plasma, Drude and Gurzhi model approaches and mean experimental
gradients \cite{18} (the sets of dots counted from top to bottom,
 respectively) are shown as the functions of separation.
The Drude parameters in the Gurzhi
model are again chosen as
(a) $\hbar\omega_p=8.68$~eV and $\hbar\gamma_{ep}=30.3$~meV and
(b) $\hbar\omega_p=9.0$~eV and $\hbar\gamma_{ep}=35$~meV.
The solid lines indicate the borders of the confidence intervals
determined in this experiment at
the 67\% confidence probability
by combining the total experimental and theoretical errors.
{}From Figs.~\ref{fg5}(a) and \ref{fg5}(b) it is seen that
both the Drude model approach and the Gurzhi model approach
are excluded by the measurement data which are consistent
with the plasma model approach to calculation of the Casimir force.

\section{Conclusions and discussion}

In the foregoing, we have discussed the extended Drude model or, as it also named,
the Gurzhi model, which describes the relaxation properties of conduction electrons
originating from electron-phonon and electron-electron scattering.
Although this model is often used in condensed matter physics and, specifically,
in the theory of high-temperature superconductors, its applications in the theory
of Casimir forces were not considered so far. Taking into account that the Casimir
puzzle remains unsolved for already 20 years (see Sec.~I), investigation of
possible extensions of the Drude model in connection with the Lifshitz theory
is a subject of much current interest.

We have considered the analytic properties of the dielectric permittivity of the
Gurzhi model. It is shown that this permittivity has a first order pole in the
upper half-plane of complex frequencies and, thus, violates the causality principle.
Additionally, within some interval along the pure imaginary frequency axis, the
Gurzhi dielectric permittivity takes the negative values.
One thus concludes that
for calculating the Casimir force it can be used only in the frequency region below
the absorption bands of a metal in combination with the dielectric permittivity
obtained from the measured optical data at higher frequencies.

Next, we have considered the imaginary part of the dielectric permittivity of the
Gurzhi model for Au at frequencies below 2~eV which can be used to calculate the
Casimir pressure by means of the Lifshitz formula. It was found to be in closer
agreement with ${\rm Im}\ve$ defined from the optical data and it leads to almost the
same, as the Drude model, extrapolation down to zero frequency,
i.e., to the region where the optical
data are not available. The concept of the Gurzhi model approach to
the Casimir force is introduced by analogy with the Drude and plasma model
approaches using the respective models combined with the optical data.
As discussed in Sec.~I,
the two latter approaches are the subject of a considerable literature in connection
with the Casimir puzzle.

The Casimir pressure between two Au plates was calculated using the Drude, plasma
and Gurzhi model approaches, as well as by using the simple Drude and plasma models,
and also by means of the Drude model applied
in the region from zero frequency to 2~eV and
supplemented by the optical data at higher frequencies. The obtained results are
compared with the data of two precise experiments on
measuring the Casimir interaction. The contribution of electron-electron
interaction to the Casimir force is estimated to be less than 0.16\%.
The Gurzhi model approach is shown to be excluded by the measurement data, as it was
demonstrated earlier for the Drude model approach. An agreement of the plasma model
approach with the measurement data at separations below $1~\mu$m is confirmed.

Although the above results do not solve the Casimir puzzle, they attach special
significance to novel experiments on measuring the Casimir force in the micrometer
separation range proposed in Refs.~\cite{41,42,43}.

%%%%%%%%%%%%%%%%%%%%%%%%%%%%%%%%%%%%%%
\section*{Acknowledgments}
The work of G.L.K.~and V.M.M.~was partially supported by the
Peter the Great Saint Petersburg Polytechnic University in the framework of the
Program ``5--100--2020".
V.M.M.~was partially funded by the Russian Foundation for Basic
Research, Grant No. 19-02-00453 A. His work was also partially
supported by the Russian Government Program of Competitive Growth
of Kazan Federal University. L.M.W.\ acknowledges financial support from
US Department of Energy under grant No.\ DE-FG02-06ER46297.
%%%%%%%%%%%%%%%%%%%%%%%%%%%%%%%%%%%%%%
%%%%%%%%%%%%%%%%%%%%%%%%%%%%%%%%%%%%%%

%%%%%%%%%%%%%
%\end{document}
\newpage
%%%%%%___Table_I___%%%%%%%%%%%%%%%%%%%
\begingroup
\squeezetable
\begin{table}
\caption{Magnitudes of the mean measured Casimir pressure \cite{11,12} (column 2)
at $T=295~$K
at different separations (column 1) are compared with the magnitudes of theoretical
pressures computed using the plasma model approach (column 3), the simple plasma
model (column 4), the Drude model approach (column 5), the Drude model used below
the first absorption band (column 6), the simple Drude model (column 7), and
the Gurzhi model approach (column 8). In all cases the Drude parameters
$\hbar\omega_p=9.0$~eV and $\hbar\gamma_{ep}=35$~meV have been used.
}
\begin{ruledtabular}
\begin{tabular}{cccccccc}
&\multicolumn{7}{c}{$|P|$~(mPa)} \\
\cline{2-8}\\
$a$~($\mu$m) & 2& 3& 4& 5& 6& 7& 8 \\
\hline \\
0.2 & $510.5\pm 1.0$ & 512.19 & 501.82 &497.35&493.80&483.29&494.57 \\
0.3 & $114.8\pm 0.5$ &115.02 &114.04 &109.69&109.14&108.13&109.25\\
0.4 &$39.2\pm 0.4$&39.15&38.98&36.70&36.57&36.39&36.59\\
0.5 &$16.8\pm 0.4$&16.81&16.76&15.50&15.45&15.41&15.46\\
0.6&$8.4\pm 0.4$&8.38&8.36&7.60&7.579&7.565&7.582\\
0.7&$4.7\pm 0.4$&4.632&4.628&4.132&4.124&4.119&4.125\\
0.8&&2.767&2.765&2.427&2.423&2.421&2.424\\
0.9&&1.754&1.753&1.512&1.5103&1.5092&1.5105\\
1.0&&1.165&1.164&0.9874&0.9864&0.9859&0.9865\\
1.1&&0.8041&0.8039&0.6699&0.6692&0.6690&0.6693\\
1.2&&0.5730&0.5729&0.4690&0.4687&0.4635&0.4687
%\hline
\end{tabular}
\end{ruledtabular}
\end{table}
\endgroup
%%%%%%%%%%%%%%%%%%%%%%%
%%%%%%%%%%%%%
%%%%%%%%%%%%%
%%%%%%%%%%%%%
%%%%%%%%%%%%%
%%%%%%%%%%%%%%%%%%%%%%%%%%
%%%%%%%%%%%%%
%%%%%%%%%%%%%

%%%%%%%%%%%%%%%%%%%%
\end{document}